\documentstyle[12pt]{article}

\begin{document}
\newcommand{\dia}{\begin{displaymath}}
\newcommand{\die}{\end{displaymath}}
\newcommand{\eqa}{\begin{equation}}
\newcommand{\eqe}{\end{equation}}
\newcommand{\eqna}{\begin{eqnarray}}
\newcommand{\eqne}{\end{eqnarray}}
\newcommand{\eqnaa}{\begin{eqnarray*}}
\newcommand{\eqnae}{\end{eqnarray*}}
\newcommand{\fraz}{\frac{1}{2}}
\newcommand{\frav}{\frac{1}{4}}
\newcommand{\frasz}{\frac{1}{\sqrt{2}}}
\newcommand{\tr}[1]{\mbox{Tr}\left\{#1\right\}}
\newcommand{\labe}[1]{\label{#1}}

\begin{titlepage}
\title{The hamiltonian formulation of QCD in terms of angle variables}
\author{Dieter Stoll\\ Department of Physics,
University of Tokyo\\ Bunkyo--ku, Tokyo 113, Japan\thanks{email:
 stoll@tkyux.phys.s.u-tokyo.ac.jp\hspace{.5cm}FAX: (81)(3)56849642}}
\date{  }
\maketitle
\begin{abstract}
For the sake of eliminating gauge variant degrees of freedom we discuss
the way to introduce angular variables in the hamiltonian formulation
of QCD. On the basis of an analysis of Gauss' law constraints
a particular choice is made for the variable transformation from gauge fields
to angular field variables. The resulting formulation is analogous to the one
of Bars in terms of corner variables. Therefore the corner or angle formulation
may constitute an useful starting point for the investigation of the low
energy properties of QCD in terms of gauge invariant degrees of freedom.
\end{abstract}

\end{titlepage}

One of the long standing problems in contemporary physics is understanding
confinement of quarks and gluons from first principles.
The difficulty in dealing with
the infrared properties of QCD is on the one hand due to the non--linear
gluonic interaction and on the other due to the constraints on the dynamics
of the fundamental degrees of freedom which originate from the requirement of
gauge invariance. In spite of the general belief that the non--linear
interaction gives rise to confinement it has been conjectured recently that
in fact the non--abelian constraints may be most important
\cite{Johnson}. Aiming at an understanding of the low energy properties
of QCD we should therefore try to develop approximations to the full QCD
dynamics after the gauge variant degrees of freedom have been identified and
isolated.

The aforementioned constraints are specified in terms of Gauss' law
operators, which generate a compact group in each point in space, telling us
that the gauge variant degrees of freedom are "angle" variables. In spite
of this observation the choice of unphysical variables is to a large extent
arbitrary due to the fact that field theory deals with an infinite number
of degrees of freedom. Therefore various decompositions into unphysical
"angle" variables and remaining physical variables are possible to arrive
at the desired separation of unphysical degrees of freedom
\cite{Johnson,Goldstone}. Although successful in that respect the variables
chosen in this way to parametrize the physical Hilbert space may be inadequate
to account in a simple way for the dynamics relevant for the low energy
properties of QCD. For further variable changes, on the other hand, the
complexity of the so derived hamiltonians constitutes a basic obstacle.

In order to avoid this problem we start from the assumption that not only
the unphysical but all variables are "angle" variables. The hamiltonian
should therefore be expressed first in terms of these angular degrees
of freedom before making a separation into gauge variant and gauge invariant
ones. To find a suitable definition of "angle" variables
in terms of gauge or electric fields we concentrate on an
analysis of Gauss' law operators. It will be shown that the form of these
operators suggests the introduction of "angles" which are non--locally
related to the gauge fields. By a variable transformation the originally
quantized gauge fields and  electric fields in the hamiltonian can be
replaced by "angle" and angular momentum operators
respectively. The resulting formulation is analogous to the one in terms of
corner variables obtained by Bars \cite{Bars}. In contrast to similar
approaches \cite{Johnson,Goldstone} the separation into
gauge variant and gauge invariant degrees of freedom is not made
from the outset in the "angle" or corner variable formulation. Therefore
it may constitute an useful starting point for the search of approximations
to the full QCD dynamics intended to understand its nonperturbative aspects.

We consider a hamiltonian formulation of SU(N) gauge theories on
a d--dimensional torus. Choosing the Weyl gauge
$A_0=0$ we have the following hamiltonian density
\dia {\cal H}=\sum_i\bar\psi(x)\gamma_i\left(i\partial_i+gA_i\right)\psi(x)
+m\bar\psi(x)\psi(x) +\fraz \sum_i E_i^a(x)E_i^a(x)+
\fraz \sum_{ij} \tr{F_{ij}F^{ij}}\ .\die
Imposing periodic boundary conditions for the gauge and anti--periodic ones for
the fermion fields we quantize canonically  ($E_i^a(x)= \partial_0A_i^a(x)$)
\eqnaa \left[E_i^a(x),A_j^b(y)\right]&=&-i\delta_{a,b} \delta_{ij}
\delta^d(x-y);\quad
a,b=1,\dots,N^2-1;\ i,j=1,\dots,d\\ \left\{\psi_{k,\alpha}^{\dag}(x),\psi_{l,
\beta}(y)\right\}&=&\delta_{k,l}\delta_{\alpha,\beta}\delta^d(x-y);\quad
k,l=1,\dots,N;\ \alpha,\beta=\mbox{Spinor indices}\eqnae
where it is understood that the $\delta$--functions are periodic, as well.
Since we have not fixed the gauge classically, Gauss' law operator is the
quantum mechanical generator of the gauge symmetry. It commutes with the
hamiltonian and therefore physical eigenstates must be invariant under
infinitesimal gauge transformations,
which implies that they must be annihilated by the generators of the symmetry
\eqna \left.G^a(x)\right|phys.> &=& 0\ \labe{constr}\\
G^a(x) &=& \sum_i\left[ \partial_iE_i^a(x)+gf^{abc}A_i^b(x)E_i^c(x) \right]
+g\psi^{\dag}(x)\frac{\lambda^a}{2}\psi(x) \label{fabc} \\
\left[G^a(x),G^b(y)\right] &=& igf^{abc}G^c(x)\delta^d(x-y) \labe{algebra}\ .
\eqne
Since the generators obey the Lie algebra of the gauge group
it is understood that out of $d\cdot (N^2-1)$ gauge degrees of freedom
only a set of $N^2-1$ "angle" variables in each point in space is changed
by gauge transformations . Consequently the constraints are satisfied if these
"angles" have been identified and Gauss' law operator has been transformed
such that it is the angular momentum operator only with respect to these
unphysical "angles". Physical states then correspond to s--wave states which
are annihilated by these angular momentum operators and the hamiltonian after
transformation  will not contain the unphysical variables anymore.

Since we want to replace gauge and electric fields by "angles" and angular
momenta in such a way that the constraints can easily be implemented, we study
the form of Gauss' law operators in detail. In 1+1 dimensions the contribution
in eq.(\ref{fabc})
\eqa f^{abc}A^b(x)E^c(x)\labe{cross}\eqe
acts locally as an angular momentum operator on $N(N-1)$ "angle variables in
either the gauge field or the elctric field representation. The missing
$(N-1)$ "angle" variables could not be identified if this was the complete
Gauss' law operator already. Therefore we must conclude that in 1+1 dimension
the full number of $(N^2-1)$ variables in each point in space can only be
eliminated due to the presence of $\partial_xE(x)$ in the Gauss' law
operators. This term not only distinguishes the gauge
fields as source of the additional unphysical variables but also
introduces a non-locality into the Gauss' law operators.
Therefore it seems natural to assume that the "angle" variables
which are unphysical are nonlocally related to the gauge field variables.
Although this argument is rigorous only in 1+1 dimensions
we assume it to be an useful hypothesis for introducing "angle" variables
in any dimensions.

An expression for the gauge fields satisfying this
requirement is\footnote{Note that throughout the paper spatial indices are
not summed over unless explicitly indicated.}
\eqna A_i(x) &=& \frac{i}{g}V_i(x) \partial_i V_i^{\dag}(x) \quad (\mbox{no
summation})\labe{afeld}\\
V_i(x) &=& P\exp\left[ig\int_0^{x_i}dz_iA_i(x_i^\perp,z_i) \right]\labe{vdef}\\
V_i(x) &=& \exp\left[i\xi_i(x)\right],\quad \xi_i(x)=\xi_i^a(x)\frac{
\lambda^a}{2};\quad 0<x_i\leq L \ \labe{xidef}\eqne
where $V_i(x)$ is a SU(N) matrix parametrized in
terms of "angles" $\xi_i^a(x)$, P denotes path ordering and $x_i^\perp$
stands for all coordinates orthogonal to $x_i$. Since this definition together
with the specific choice of paths in eq.(\ref{vdef}) leads to an unique
relation\footnote{With the choice \protect $0<x_i\leq L$ the
"angles" are uniquely determined from gauge fields by eq.(\ref{vdef}) if the
derivative is taken from one side only \protect{
$\partial_if(x_i) = \lim_{\epsilon\rightarrow 0} \frac{1}{\epsilon}\left[
f(x_i)-f(x_i-\epsilon)\right] ;\quad 0<x_i\leq L \labe{deriv}$}.
In this way it is possible to work with periodic, although not continuous
"angles" and angular momenta.} between $\xi_i(x)$ and $A_i(x)$ a
change of variables from $A_i(x), E_i(x)$ to $\xi_i(x)$ and the corresponding
angular momenta $J_i(x)$ becomes feasible. Using eq.(\ref{afeld})
we rewrite fermionic and magnetic part of the hamiltonian
\eqa \bar\psi(x)\left\{\gamma_i\left[i\partial_i+gA_i(x)\right] +m\right\}
\psi(x)= \left[\bar\psi(x)U_i(x)\right] \left\{ \gamma_ii\partial_i +m\right\}
\left[U_i^{\dag}(x) \psi(x)\right] \labe{ferm}\ ,\eqe
\eqna \left[D_i,D_j\right] &=& \left[i\partial_i+gA_i,i\partial_j+gA_j\right] =
-U_i\left\{ \partial_i\left[\left(U_i^{\dag}U_j\right)\partial_j
\left(U_j^{\dag}U_i \right)\right] \right\}U_i^{\dag}\ , \\ \Rightarrow
\tr{F_{ij}F_{ij}} &=& \frac{-1}{g^2}\tr{\partial_i\left[\left[
\left(U_i^{\dag}U_j\right) \partial_j\left(U_j^{\dag}U_i\right)\right]\right]
\left[\partial_i\left[\left( U_i^{\dag}  U_j\right)\partial_j
\left(U_j^{\dag}U_i\right)\right]\right]^{\dag}}\nonumber  . \eqne
In order to reformulate the electric part of the hamiltonian which contains the
conjugate momenta of the gauge fields, we
introduce the angular momentum operators $J_k^c(z)$, the definition
of which may be found in the appendix eq.(\ref{jdef}). These operators
generate translations in the space of "angles" $\xi_k$ as may be
seen from the commutation relations
\eqna \left[J_i^a(x),V_j(y)\right]&=&\delta_{i,j}\delta^d(x-y)V_j(y)
\frac{\lambda^a}{2}\labe{jvcom}\\   \left[J_i^a(x),J_j^b(y)
\right]&=&\delta_{i,j}if^{abc}J_i^c(x)\delta^d(x-y)\ .\nonumber \eqne
We note that due to the periodicity of $V_i,J_i$ the $\delta$--functions
in these expressions are periodic, as well. Introducing furthermore the
orthogonal matrices $N_i$
\eqna N_i^{ac}(x) &=& \tr{V_i^{\dag}(x)\frac{\lambda^a}{2}V_i(x)\lambda^c}
\labe{ndef} \\  \left[J_i^b(z),N_j^{ac}(x)
\right]&=& if^{bce}N_j^{ae}(x)\delta_{i,j}\delta^d(x-z) \nonumber \eqne
and using the identity eq.(\ref{avder}) stated in the appendix, we
find for the electric part of the hamiltonian density the expression
\eqna E_i^a(x) &=& g\int d^dz \delta^{d-1}(z_i^\perp-x_i^\perp)\theta(z_i-x_i)
\theta(x_i)N_i^{ac}(x) J_i^c(z)\labe{edef} \nonumber  \\
\fraz E_i^a(x)E_i^a(x) &=& \frac{g^2}{2}\int_{x_i}^Ldz_iJ_i^b(
x_i^\perp,z_i)\int_{x_i}^Ldz_i^\prime J_i^b(x_i^\perp,z_i^\prime)\ .\labe{hej}
 \eqne
Collecting all the results the hamiltonian density reads
\eqna {\cal H}&=&\sum_i\left[\bar\psi(x)U_i(x)\right] \left[ \gamma_i
i\partial_i +m \right] \left[ U_i^{\dag}(x)\psi(x)\right]
\nonumber \\ & +& \frac{g^2}{2}
\sum_i\int_{x_i}^Ldz_i J_i^b(x_i^\perp,z_i)\int_{x_i}^Ldz_i^\prime
J_i^b(x_i^\perp,z_i^\prime) \labe{hneu}\\
&+& \frac{1}{2g^2}\sum_{ij} \tr{\left[\partial_i\left[\left(
U_i^{\dag}U_j\right)\partial_j\left(U_j^{\dag}U_i\right)\right]\right]\left[
\partial_i\left[\left(U_i^{\dag}U_j\right)
\partial_j\left(U_j^{\dag}U_i\right)\right]\right]^{\dag}}\nonumber \eqne
which is the "angle" representation we have been looking for. Note that
the locality of the hamiltonian has been lost although we have not been
fixing the gauge yet. The electric part of the hamiltonian is non--local
and it shows already the linear "potential" $|z_i-z_i^\prime|$ characteristic
for both the axial gauge formulation and the strong coupling limit in
lattice gauge theory. We observe also that
the hamiltonian in QED, corresponding to eq.(\ref{hneu}), is obtained by
dropping the summations over color indices which shows the similarity
of abelian and non--abelian gauge theories in the "angle" formulation.\\
Finally we want to consider the form of Gauss' law operator in these
variables. We find
\eqna gf^{abc}A_i^b(x)E_i^c(x) &=& -g\left[\partial_iN_i^{ad}(x)\right]
\int dz_i\theta(z_i-x_i)\theta(x_i) J_i^d(x_i^\perp,z_i)\\
\partial_iE_i^a(x) &=& g\left[\partial_iN_i^{ad}(x)\right]
\int dz_i\theta(z_i-x_i)\theta(x_i)J_i^d(x_i^\perp,z_i)\nonumber \\
 && +gN_i^{ad}(x)\int dz_iJ_i^d(x_i^\perp,z_i)\partial_i\left[\theta(z_i-x_i)
\theta(x_i)\right]\eqne
and taking the sum of these two contributions and the charge density operator
we obtain for the Gauss' law operators the following expression
\eqa G^a(x) = -g\sum_i\left[N_i^{ad}(x)J_i^d(x)-\delta(x_i)\int dz_i
J_i^a(x_i^\perp,z_i)\right] + g\psi^{\dag}(x)\frac{\lambda^a}{2}
\psi(x)\nonumber\ . \eqe
Thus we have arrived at a continuum formulation of non--abelian gauge theories
entirely in terms of angular degrees of freedom. The objective for doing so was
the wish to have a formulation that leaves us freedom in choosing appropriate
unphysical "angle" variables. Since the introduction of appropriate coordinates
is already very important in quantum mechanics, this freedom may be crucial
for developing useful
approximations to understand the low energy properties of QCD. The formulation
we found can be shown to be equivalent in a finite volume to Bars corner
variable formulation. Therefore also the obvious similarity, in terms of
variables, to the lattice hamiltonian approach to QCD \cite{Kogut} can be
made precise by appropriately discretizing the spatial variables in our
results \cite{Bars1}. Thus the "angle" variable formulation of QCD in the
continuum  shows a close relationship with the lattice QCD approach and it has
the built--in freedom in selecting unphysical variables. Moreover for the
axial gauge representation of QCD \cite{Lenz} it has been shown
already that this reformulation leads to great technical simplifications in
the elimination of unphysical variables \cite{Stoll}. All these
advantages together may render the "angle" formulation an useful starting
point for investigating non--perturbative aspects of QCD in terms of gauge
invariant degrees of freedom.\\
\begin{center} {\large \bf Acknowledgements}\end{center}
We would like to thank Profs. K.Ohta, M.Thies and K.Yazaki for helpful
discussions. Financial support by the "Japan Society for the Promotion of
Science" is furthermore gratefully acknowledged.
\vspace{.3cm}\\
{\large\bf Appendix}\\
{}From the definition of the angles $\xi_i(x)$ eq.(\ref{xidef}) it is
clear that the $\xi_i$ only depend on the gauge
fields $A_i$. Infinitesimal changes of the matrix $V_i$ are related to
changes in the "angles" $\xi_i$ through
\eqna V_i^{\dag}(x)\delta V_i(x) &=& iM_i^{ab}(x)\frac{\lambda^b}{2}\delta
\xi_i^a(x)\\
\delta\xi_i^b(x)&=&-iW_i^{cb}(x)\tr{V_i^{\dag}(x)\delta V_i(x)\lambda^c}
\\ \frac{\delta\xi_i^b(z)}{\delta A_j^a(x)}&=&\tr{V_i^{\dag}(z)\frac{-i\delta}{
\delta A_j^a(x)}V_i(z)\lambda^c}W_i^{cb}(z)\eqne
with $W_i(x)=M_i(x)^{-1}$. The angular momentum operators are then defined by
\eqa J_k^c(z)=W_k^{cb}(z)\frac{-i\delta}{\delta\xi_k^b(z)}
\labe{jdef}\eqe
and have commutation relations given in eqs.(\ref{jvcom}).

To obtain the representation eq.(\ref{edef}) for the electric field
operator in terms of these "angular" momentum operators we have to
make use of the following identity for the derivative of the path
ordered integral $V_i$ with respect to $A_j$
\eqa \frac{-i\delta V_i(z)}{\delta A_j^a(x)}=g\delta_{i,j}\delta^{d-1}(
z_i^\perp-x_i^\perp)\theta(z_i-x_i)\theta(x_i)V_i(z)
V_i^{\dag}(x)\frac{\lambda^a}{2}V_i(x)\labe{avder}\eqe
which may be verified by taking the derivative with respect
to $z_i$ \cite{Lenz}.

\end{document}